# Arbitrary access temporal pulse cloaking and restoring in periodically poled lithium niobate

Yi'an Liu, Yuping Chen*, Licheng Ge, Haowei Jiang, Guangzhen Li, Xianfeng Chen*

State Key Laboratory of Advanced Optical Communication Systems and Networks,
Department of Physics and Astronomy, Shanghai Jiao Tong University,
800 Dongchuan Road, Shanghai 200240, China

Temporal cloaks have inspired the innovation of research on security and efficiency of quantum and fiber communications for concealing temporal events. The existing temporal cloaking approaches possessing ps ~ns cloaking windows employed the third-order nonlinearity mostly. Here we explore a temporal cloak for perpetually concealing pulse events using high efficiency second-order nonlinearity. A temporal pulse event was cloaked as a continuous wave with constant intensity and restored by polarization dependent sum frequency generation processes between the continuous wave and probe. The variety of quasi phase-matching second-order nonlinear processes in periodically poled lithium niobate make sure the pulse event cannot be filched during transmission. The proposed temporal cloak predicts hopeful applications in information security of photonic integrated circuit on lithium niobite thin film.

## 1. INTRODUCTION

Security and efficiency are eternal topics in the development of optical communications. The dramatical improvement in high-speed optical communication supports large-capacity data transmission [1–3], which urgently calls for reliable methods to ensure the security in ever-increasing data transmission. Therefore, temporal cloaking emerges as an effective tool for secure communications. The idea of temporal cloaking [4] is analogous to spatial cloaking [5–7]. The original temporal cloaking worked in a time gap where events were concealed. With a time gap created by time lens and dispersive medium, temporal cloaking has been firstly demonstrated in optical fibers [8]. However, this approach suffered from high energy consumption and low efficiency. Then temporal cloaking at telecommunication data rate was demonstrated with periodic time gap obtained by fractional temporal Talbot effect [9, 10]. Where this method worked on periodical events and the cloaking window was about 196ps. Temporal cloaking in fiber [11] and silicon microring [12] recently improved the cloaking window from picosecond-level to nanosecond-level. The principle of all temporal cloaking above was attaining mismatch between probe wave and events in the time-spatial domain, so events could only be concealed in a finite time window. From another point of view, by controlling the interaction (four-wave mixing) between probe wave and events in fiber, a perpetual temporal cloaking was also obtained [13]. The limitation is that cloaking would fail easily when the spy switches the state of polarization (SOP) of probe wave, because either vertical or horizontal polarized probe wave matches the four-wave mixing.

In contrast with the current schemes worked on finite cloaking time windows employing the material third-order nonlinearity [8–13], the lithium niobate (LN) can perform a perpetual temporal cloaking based on its large second-order nonlinearity and acts vital roles in modern optical communication [14–17]. Recently many exciting works in LN photonic integrated circuit (PIC) [18–21] and on-chip quasi-phase-matching (QPM) nonlinear frequency conversion [22, 23] are motivated in the background that Moore’s law [24] is approaching a limit, and also benefit from the commercialization of lithium niobate on insulator (LNOI) [25]. Therefore, the demonstrated temporal cloaking scheme in this paper will be explored to be integrated into periodically poled lithium niobate on insulator (PPLNOI) [26].

Owing to the wide transparent window (0.4-5.0μm) and its rich nonlinearity in LN, we design and demonstrate an infinite time gap temporal cloaking at C-band in periodically poled lithium niobite (PPLN). The pulse (events) are hidden in an intensity-constant continuous wave (CW) with preset SOP, modulated by a LN polarization modulator. The pulse can later be restored by controlling the QPM sum-frequency generation (SFG) in PPLN [27] between a probe wave and the CW. The key to this temporal pulse cloaking is the designed QPM periods of PPLN. In consideration of the uncertainty of the QPM condition of different SFG, only using the PPLN with a certain period can the pulse be restored. Furthermore, protocols based on four modes of SFG have been designed to ensure the reliability of the cloaking in our experiment.

## 2. PRINCIPLE AND METHODS

The schematic of temporal pulse cloaking is depicted in Figure 1. A sender intended to transmit an optical pulse to a user. Normally, the sender sent the pulse in a public channel. However, a spy could filch the pulse easily, as shown in Figure 1 a. In order to cloak the pulse from the spy, it should be handled before transmission, which is shown in Figure 1 b. The sender firstly chose a random SOP sequence for a probe wave used in the following SFG and shared the sequence with the user in a public channel. Then the sender detected the intensity of the pulse and input an electric square wave to a polarization switch. The switch changed a linear polarized CW into varying orthogonal polarized one, according to the electric square wave. Because the sender possessed the period of PPLN belonging to the user, he accurately arranged the SOP of the CW, according to the corresponding QPM condition in that PPLN sample, and delivered the CW to the user in a public channel. The user received the CW and implemented SFG with a properly polarized probe wave in his PPLN, and restored the pulse by detecting the intensity of idler wave. As for the spy, he filched the CW and the SOP of the probe wave, nevertheless the SOP was not related to the transmitted pulse. Therefore, the pulse was cloaked in the transmission process and restored by the user whenever the SFG was implemented.

The QPM SFG in PPLN obeys energy conservation and phase matching condition, which are expressed as

$$\frac{1}{n_1\lambda_1}+\frac{1}{n_2\lambda_2}=\frac{1}{n_3\lambda_3} \qquad (1)$$

$$\Delta K=\frac{2\pi n_1}{\lambda_1}+\frac{2\pi n_2}{\lambda_2}-\frac{2\pi n_3}{\lambda_3}-\frac{2\pi}{\Lambda} \qquad (2)$$

$l_1$, $l_2$ and $l_3$ represent the wavelength of CW, probe and idler wave respectively and $L$ is the period of PPLN. For the user, $l_1$ and $L$ are given while $l_2$ and $l_3$ are unknown, so $l_2$ could be derived using (1) and (2). For the spy, however, only $l_1$ can be obtained while $l_2$, $l_3$ and $L$ are unknown. So the spy is not able to solve the above equations containing four variables with only one of them confirmed. In this case, the spy can’t imitate the SFG process physically for filching the pulse.

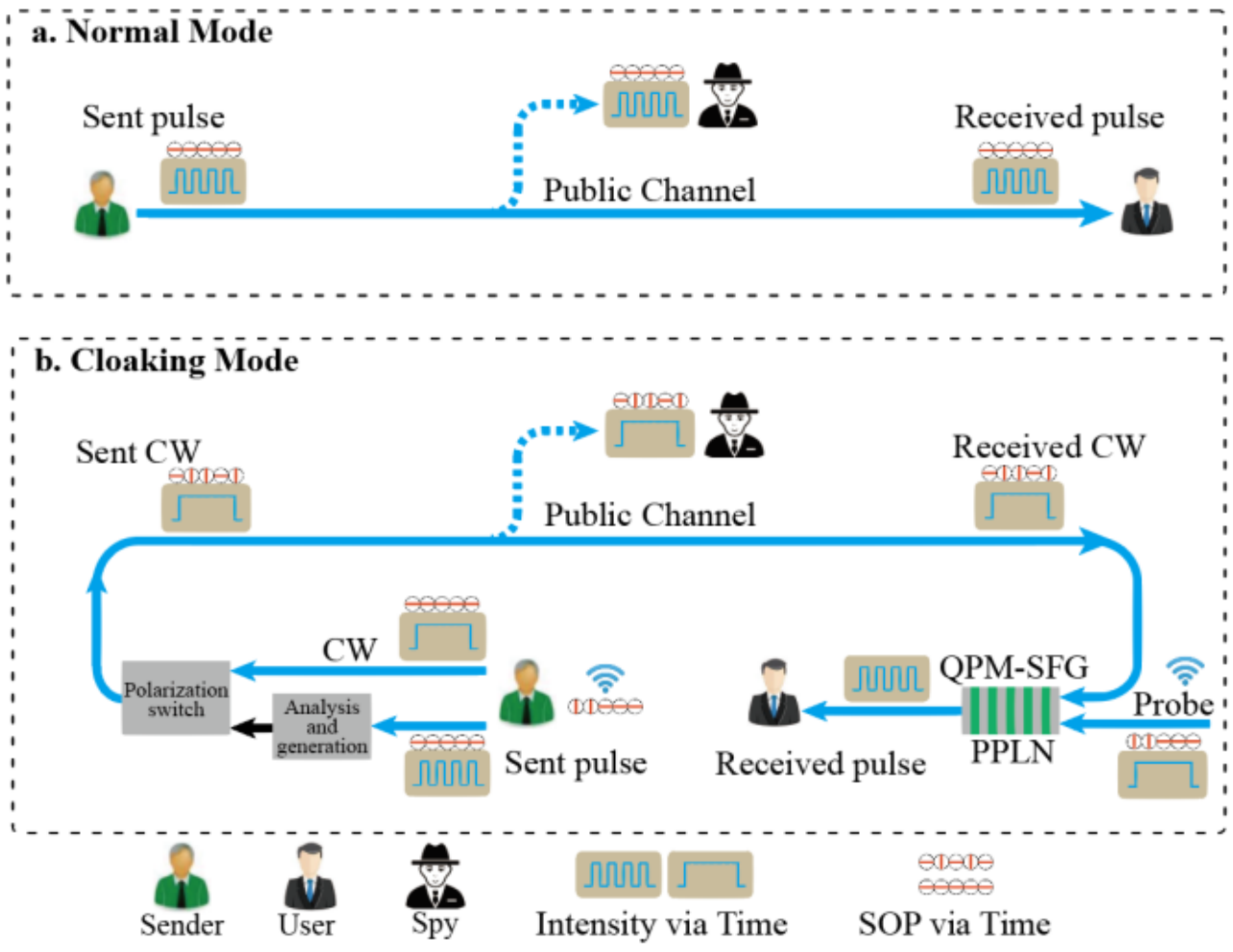

Fig. 1. Schematic of temporal pulse cloaking. a. The sender delivered a pulse to the user in a public channel. b. The pulse delivered in a public channel was cloaked. The blue arrow line denotes optical wave and the black one denotes electric signal. QPM SFG: quasi-phase-matching sum-frequency generation.

According to the nonlinear susceptibility tensor of LN [28], the second-order nonlinear optical susceptibility of three-wave mixing can be expressed as

$$\begin{aligned} &P_y(\omega_3) \propto d_{22}E_y(\omega_1)E_y(\omega_2) + d_{31}E_y(\omega_1)E_z(\omega_2) \\ &+d_{31}E_z(\omega_1)E_y(\omega_2) \\ &P_z(\omega_3) \propto d_{31}E_y(\omega_1)E_y(\omega_2) + d_{33}E_z(\omega_1)E_z(\omega_2) \end{aligned} \tag{3}$$

Assume three optical waves propagate along x-axis and the crystal axis of LN is z-axis. $E_y$ and $E_z$ will represent ordinary (*o*) and extraordinary (*e*) light respectively. We can find from (3) that there are five kinds of SFG (*e* + *e* → *e*, *o* + *o* → *e*, *o* + *o* → *o*, *o* + *e* → *o*, *e* + *o* → *o*). In the temporal cloaking process, the SOP of idler wave has no difference to the result, so there are four modes of SFG (*e* + *e* → 1, *o* + *o* → 1, *e* + *o* → 1, *o* + *e* → 1) in principle. The number 1 depicts the idler wave is generated in SFG and its intensity is detected by photoelectric detector (PD) as 1 (binary system). Four protocols of temporal cloaking based on SFG are shown in Table 1. For example, in Table 1 a, the protocol SFG ($Probe^e + CW^e \rightarrow Idler^1$) means that if the probe wave is *e* and the CW is *e*, the idler wave is detected. No matter which protocol is employed for the temporal cloaking process, the SOP of the CW in Figure 1 has no correlation to the intensity of the idler wave (the pulse). As is explained previously that the spy is not able to filch the pulse physically, here we find that the spy can't filch the pulse by analyzing the correlation between the SOP of the CW and the pulse. Suppose the spy adopted one of the four protocols and analyzed the SOP of the probe wave and CW jointly, the probability of filching success is (1/4). When the pulse was transmitted using all of the four protocols with a certain order that the sender and the user preset, it would bring more difficulties to the spy. Let the pulse be cut into *n* pieces, the probability of filching the pulse correctly and completely is $(1/4)^n$. Therefore, the reliability of the temporal cloaking process is sufficient finally.

**a. Protocol I** SFG($\text{Probe}^e + \text{CW}^e \rightarrow \text{Idler}^1$)

| | | | | | | | | | | |
|---|---|---|---|---|---|---|---|---|---|---|
| Probe(SOP) | *o* | *e* | *o* | *e* | *o* | *o* | *e* | *e* | *o* | *o* |
| CW(SOP) | *o* | *o* | *e* | *e* | *e* | *o* | *e* | *o* | *o* | *e* |
| Idler(Intensity) | 0 | 0 | 0 | 1 | 0 | 0 | 1 | 0 | 0 | 0 |

**b. Protocol II** SFG($\text{Probe}^o + \text{CW}^o \rightarrow \text{Idler}^1$)

| | | | | | | | | | | |
|---|---|---|---|---|---|---|---|---|---|---|
| Probe(SOP) | *o* | *e* | *o* | *e* | *o* | *o* | *e* | *e* | *o* | *o* |
| CW(SOP) | *o* | *o* | *e* | *e* | *e* | *o* | *e* | *o* | *o* | *e* |
| Idler(Intensity) | 1 | 0 | 0 | 0 | 0 | 1 | 0 | 0 | 1 | 0 |

**c. Protocol III** SFG($\text{Probe}^e + \text{CW}^o \rightarrow \text{Idler}^1$)

| | | | | | | | | | | |
|---|---|---|---|---|---|---|---|---|---|---|
| Probe(SOP) | *o* | *e* | *o* | *e* | *o* | *o* | *e* | *e* | *o* | *o* |
| CW(SOP) | *o* | *o* | *e* | *e* | *e* | *o* | *e* | *o* | *o* | *e* |
| Idler(Intensity) | 0 | 1 | 0 | 0 | 0 | 0 | 0 | 1 | 0 | 0 |

**d. Protocol IV** SFG($\text{Probe}^o + \text{CW}^e \rightarrow \text{Idler}^1$)

| | | | | | | | | | | |
|---|---|---|---|---|---|---|---|---|---|---|
| Probe(SOP) | *o* | *e* | *o* | *e* | *o* | *o* | *e* | *e* | *o* | *o* |
| CW(SOP) | *o* | *o* | *e* | *e* | *e* | *o* | *e* | *o* | *o* | *e* |
| Idler(Intensity) | 0 | 0 | 1 | 0 | 1 | 0 | 0 | 0 | 0 | 1 |

Table 1. Four protocols of temporal pulse cloaking based on SFG between the CW (sent in a public channel) and the probe wave (imported by the user). *o* and *e* represent ordinary and extraordinary light respectively. 1 and 0 depict whether the idler wave of SFG is generated.

## 3. SIMULATION AND EXPERIMENT

We simulate each mode of SFG used in the temporal pulse cloaking. Figure 2 (a) is SFG ($e + e \rightarrow 1$) (mode I) and the period of PPLN is 19.6*m*m, as well as the temperature is 50 °C. The bright-line is related to certain wavelengths of the probe wave and CW chosen by QPM condition. Figure 2 (b) is SFG ($o + o \rightarrow 1$) (mode II) on condition that the period of PPLN is 20.3*m*m and the temperature is 25.1 °C. In this case, the proper wavelength of the probe wave and CW can be chosen in a broad region, which refers to broadband QPM SFG. PPLN with 700*m*m period is needed for the realization of SFG ($e + o \rightarrow 1$) and SFG ($o + e \rightarrow 1$) in c-band after some numerical simulation. From the result of the simulation, we find that mode I and II SFG are convenient to be realized in our experiment, mainly because the high-efficiency SFG can be obtained in PPLN with a proper period at room temperature. Figure 2 (c) shows the experiment of mode I SFG. The highest SFG conversion frequency is located at 777.5nm as the idler wave, where the probe wave is at 1554.1nm with 9.94mW, and the CW is at 1556.2nm with 10.32mW. Figure 2 (d) shows the mode II SFG, where the wavelength of the CW and probe wave are at 1563nm and 1553.8nm respectively, both with input power at 50mW. And the dashed line shows the simulation of the broadband SFG [29] calculated with the Sellmeier equation [30].

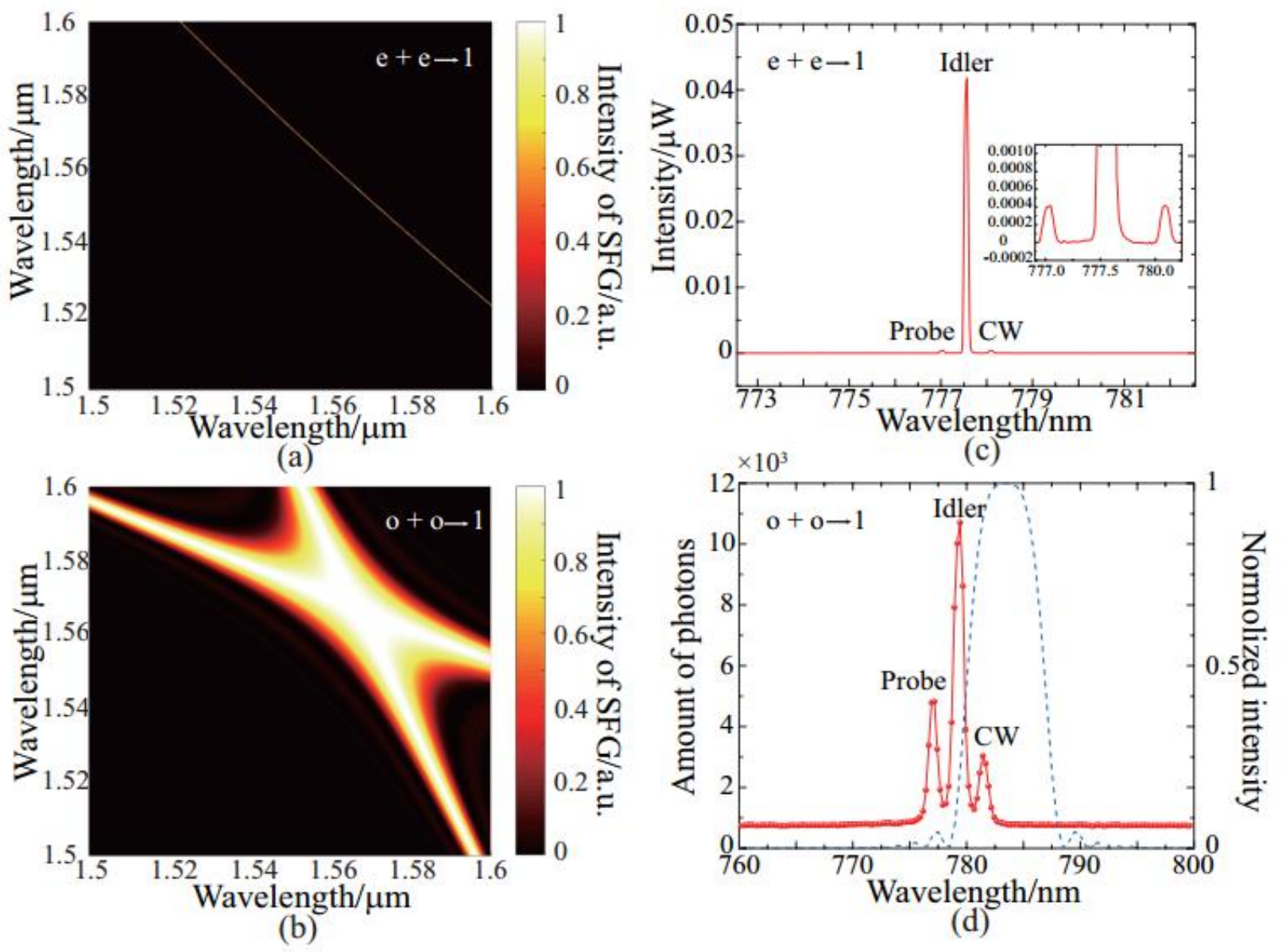

Fig. 2. The simulation (a, b) and experiment (c, d) results for SFG ($e + e \rightarrow 1$) and ($o + o \rightarrow 1$)

We choose protocol I and II in Table 1 to deliver the pulse in the cloaking experiment. The setup, shown in Figure 3, consists of three parts, the sender, user and spy. The CW is produced by a tunable laser at 1563nm from the sender's side. Meanwhile, an electric square wave is loaded into a polarization modulator by a function generator, according to the pulse we want to deliver. The SOP of CW rotates in the polarization modulator at the time window when the electric square wave has non-zero intensity. With the half-wave voltage set at these time windows, the linear polarized CW gets serial orthogonal SOP, and the sender finishes cloaking the pulse. The probe wave is given by another tunable laser at 1553.8nm from the user's side and is enlarged by an erbium-doped fiber amplifier(EDFA) to ensure the power of the idler wave. The user detects the intensity of the idler wave with a PD. The spy uses a 10:90 coupler to filch the pulse and detects its intensity by a PD. The result of detection is shown in Figure 4. The sender cloaks a pulse in a CW with alternatively *o* and *e* SOP, then the user imports a probe with preset *e* (Figure 4.a) or *o* (Figure 4.b) SOP for SFG, and detects a waveform of the intensity that matches with the pulse. The non-return-to-zero intensity results from a small quantity of SHG of the CW and probe wave. The spy detects the CW sent in the public channel and fails to filch the pulse. It shows that the pulse is successfully delivered from the sender to the user while cloaked under the spy's filching.

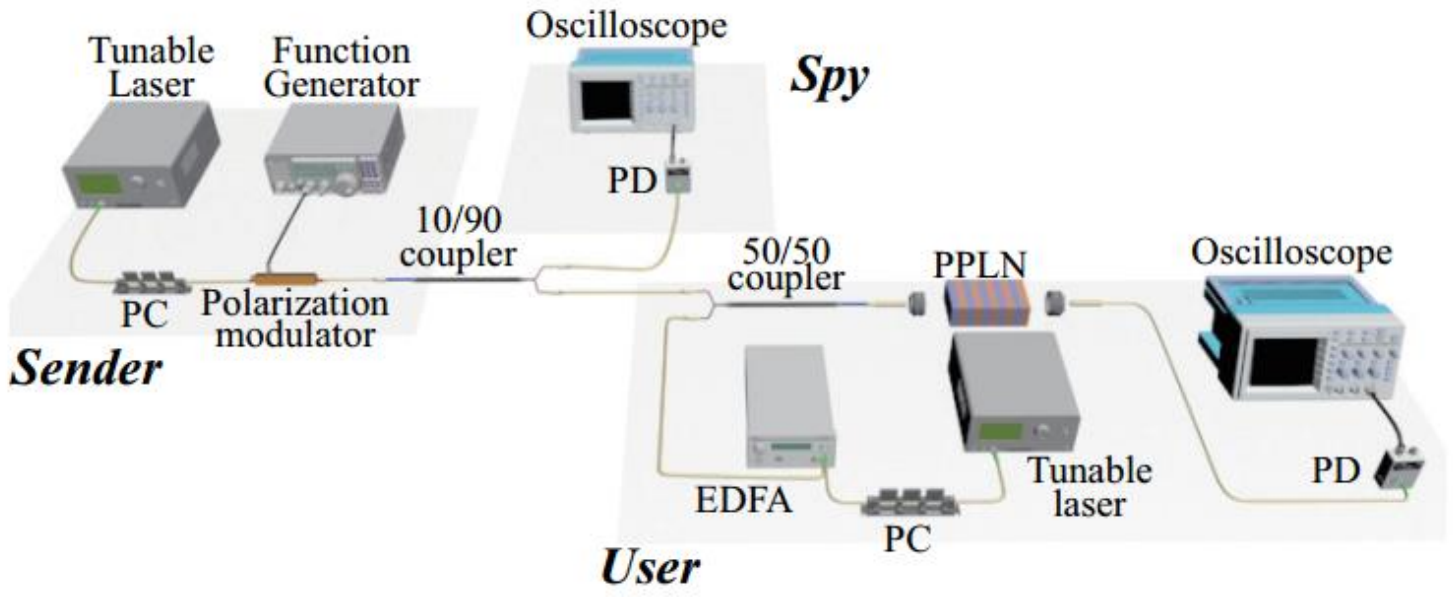

Fig. 3. The experimental setup. The sender, user, and spy are labeled with three shaded squares under the instruments. PC: polarization controller, PD: photoelectric detector, EDFA: erbium-doped fiber amplifier.

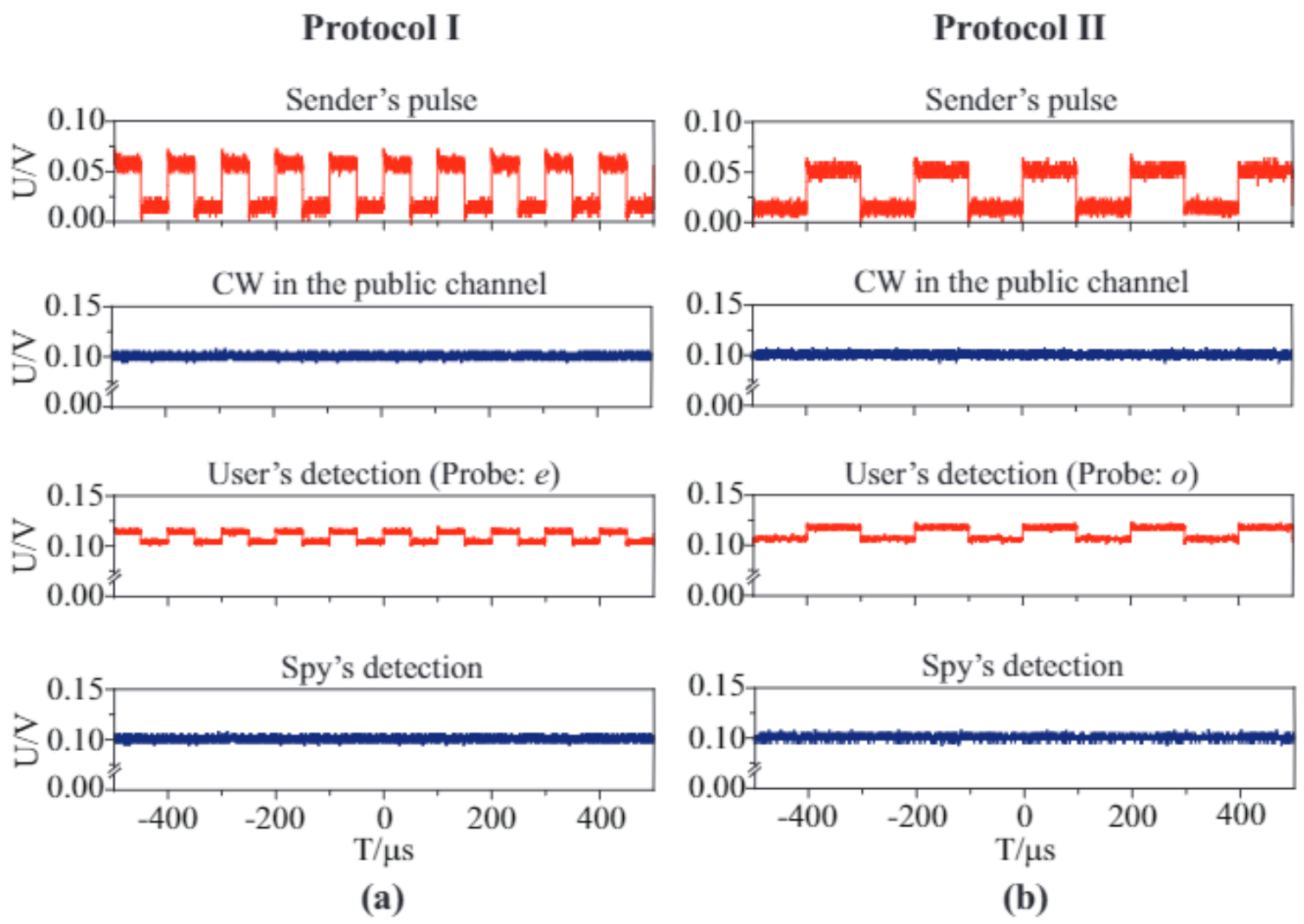

Fig. 4. The result of detection. The cloaking protocols used in the experiment are the protocol I (a) and II (b). For convenience, the CW sent in the public has alternatively *o* and *e* SOP and the SOP of the probe wave is *e* or *o* all the time.

## 4. CONCLUSION AND DISCUSSION

In conclusion, we present a scheme for temporal pulse cloaking and successfully deliver pulse events privately. It is significant we can restore the cloaked pulse events as we like at any time. We also estimate the reliability of this temporal cloaking process. The critical point of this work is introducing the principle of temporal pulse cloaking based on the rich second-order nonlinearity in PPLN. Especially with LNOI and PPLN commercialization and the fabrication technique progress for low-loss LN waveguide recently, we can design and fabricate low-loss waveguide on PPLNOI to realize high-efficiency QPM SFG in communication band for the completeness of on-chip perpetual temporal pulse cloaking. We hope this temporal cloaking has extensive application in the field of on-chip communication and information security.

## FUNDING INFORMATION

National Key R&D Program of China (2017YFA0303700); National Natural Science Foundation of China (NSFC) (11574208 and 91950107).